# Enhancement of the Monolayer WS$_2$ Exciton Photoluminescence with a 2D-Material/Air/GaP In-Plane Microcavity

Oliver Mey[1], Franziska Wall[1], Lorenz Maximilian Schneider[1], Darius Günder[1], Frederik Walla[2], Amin Soltani[2], Hartmut G. Roskos[2], Ni Yao[3], Peng Qing[3], Wei Fang[3], and Arash Rahimi-Iman*[1]

1) Department of Physics and Materials Sciences Center, Philipps-Universität Marburg, Marburg, 35032, Germany
2) Physikalisches Institut, Johann Wolfgang Goethe-Universität, 60438 Frankfurt am Main, Germany
3) State Key Laboratory of Modern Optical Instrumentation, College of Optical Science and Engineering, Zhejiang University, Hangzhou 310027, China

**Abstract:**

Light-matter interaction with two-dimensional materials gained significant attention in recent years leading to the reporting of weak and strong coupling regimes, and effective nano-laser operation with various structures. Particularly, future applications involving monolayer materials in waveguide-coupled on-chip integrated circuitry and valleytronic nanophotonics require controlling, directing and optimizing photoluminescence. In this context, photoluminescence enhancement from monolayer transition-metal dichalcogenides on patterned semiconducting substrates becomes attractive. It is demonstrated in our work using focussed-ion-beam-etched GaP and monolayer WS$_2$ suspended on hexagonal-BN buffer sheets. We present a unique optical microcavity approach capable of both efficient in-plane and out-of-plane confinement of light, which results in a WS$_2$ photoluminescence enhancement by a factor of 10 compared to the unstructured substrate at room temperature. The key concept is the combination of interference effects in both the horizontal direction using a bull's-eye-shaped circular Bragg grating and in vertical direction by means of a multiple reflection model with optimized etch depth of circular air-GaP structures for maximum constructive interference effects of the applied pump and expected emission light.

## Introduction: Enhancing light–2D-matter interaction

Transition-metal dichalcogenides (TMDCs) in the monolayer regime have been extensively studied as two-dimensional (2D) semiconductors due to their extraordinary strong light-matter interaction [1,2]. When thinned down to a single crystal layer with sub-nanometer thickness, the indirect-to-direct energy gap transition gives rise to pronounced photoluminescence (PL) [3]. Many optoelectronic devices have been envisioned to exploit these special optoelectronic properties and the 2D nature of TMDCs and their heterostructures, such as novel photo-detection and light-emitting devices, including nanolasers with ultra-high efficiency, or light-matter-coupled systems for cavity quantum-electrodynamics studies (see Refs. [4–6] and Refs. within). In addition, the circular dichroism due to spin-valley locking motivates the employment in future "valleytronic" applications [7]. A very promising material for nanophotonic applications in this group is WS$_2$, as it is comparatively easy to exfoliate, features a distinct and bright A-exciton luminescence peak [8]—the brightest and narrowest mode among its family members—and a high quantum yield (~6% reported [9]).

Integrating 2D-layered structures into optical, electrical or optoelectronic circuits to effectively perform operations in microchips still remains a technological as well as a design challenge, which is tackled from many sides by improving growth and structure fabrication as well as by developing suitable photonic platforms for efficient light-matter interaction. Efforts have been undertaken to incorporate monolayers into microcavities and photonic structures of different kind for lasing studies [9–11], weak [12–16] or even strong light-matter coupling [17–20]. Furthermore, it has been shown, that, by optimizing laser light in-coupling into and PL out-coupling from a monolayer flake, the PL intensity can be strongly modified [21,22].

The aim of this work is to enhance both light-matter interaction and the obtainable PL of a monolayered system—using $WS_2$ as suitable representative of its class—by means of an in-plane-microcavity as well as a thin-film-interference via substrate pattering.

To obtain higher PL yield for monolayer materials, two pathways of structure design for 2D-materials substrates have been pursued and combined.

Firstly, we consider enhancing light in- and out-coupling conditions at the position of the monolayer $WS_2$ flake, by means of designing a local field enhancement in the out-of-plane direction. The internal reflections in a $WS_2$-air-substrate vertical air-gap cavity are tuned to provide constructive interference at the monolayer position, and, at the same time, the out-coupling efficiency of the monolayer to the air is maximized. This is achieved using an optimized depth of the underlying trench in the substrate. Here, a high vertical reflectivity for light is preferable, which can be achieved by using a substrate with high refractive index for a good index contrast to air.

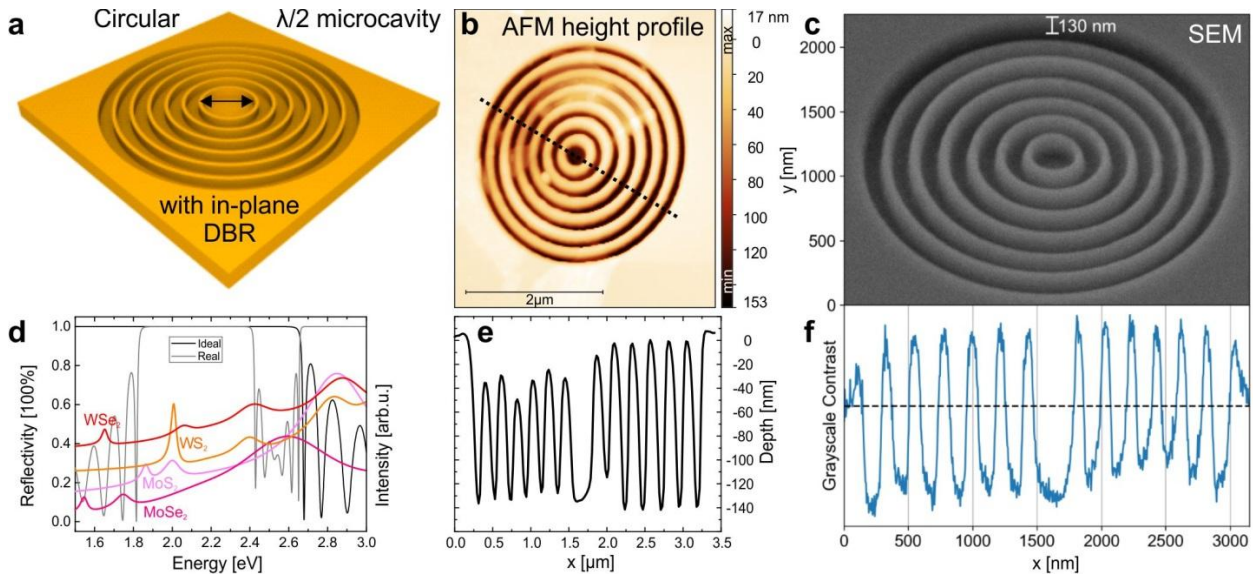

**Figure 1 | Design and fabrication of a radial-symmetric in-plane optical microcavity.** *(a) Schematic representation of a "bull's-eye"-shaped circular in-plane distributed-Bragg-reflector(DBR)-based optical microcavity (short CIDBROM) as a model system for in-plane confinement scenarios in connection with 2D-layered material systems. (b) atomic-force microscopy (AFM) and (c) scanning-electron microscopy (SEM) image of a FIB-produced CIDBROM. The depth of 130 nm is determined by an AFM measurement. The cross-sectional height profile (e) and gray-scale contrast line (e), respectively, are depicted below the corresponding image. (d) Transfer-matrix-calculated stop band of a 1D air-Bragg grating without a central cavity defect in a GaP environment (in-plane reflection spectrum, gray line, left axis) for the focussed-ion-beam-produced structure shown in (c). Here, the typical exciton absorption peaks of*

*common TMDCs (colour lines with labels, right axis) are inserted with intensities comparable to each other and with constant offset for clarity (data after [23]). The spectrum of the targeted design (black solid line) with center wavelength corresponding to the A-exciton line of WS$_2$ is shown for the ideal dimensions of air (154 nm) and GaP (46 nm) resembling λ-quarter layers.*

Secondly, we aim at enhancing the electric field in a spatially well-defined microscopic area by confinement effects which can be achieved by producing an in-plane optical microcavity with a lateral Bragg mirror structure (see **Fig. 1**). This design allows for the in-plane component of the PL to be efficiently confined to the central position of such lateral microcavity (here with a radial-symmetric "bull's-eye" structure also studied for different applications as well as wavelength regions [24–26]). The cavity permits the light field to resonantly interact with the emitter, promising a Purcell effect [27]. Furthermore, a directed emission that is favorable in most applications can be achieved [24]. Here, a higher substrate refractive index is favorable as well, since the number of in-plane distributed-Bragg-reflector (DBR) layers leading to the overall near-total reflectance and photonic-stop-band formation can be accordingly reduced with a view towards a reasonably high quality factor (see "air-Bragg" microcavities [28]).

In this work, we present a bull's-eye-shaped circular in-plane optical microcavity with DBR ring structure (short CIDBROM) prepared by focussed-ion-beam (FIB) milling of a GaP substrate, as shown in **Fig. 1**. The radial symmetry was chosen for a maximized PL enhancement and out-of-plane collection, while linear resonators with opposing parallel in-plane DBRs (PIDBROMs) are used for waveguide-based optical integrated circuitry, e.g. with strips attached to the lower-reflectance side. Here, few-layer hBN-buffered monolayer WS$_2$ on top of patterned GaP is studied at room temperature optically, using micro-reflection-contrast as well as micro-PL measurements and scattering-type near-field microscopy. The PL measurement results obtained from the CIDBROM-supported 2D-material stack are compared to corresponding data from a circular reference structure with no grating in order to show the structural influence. On the other hand, the sample is normalized by the PL of the 2D-material stack on a GaP substrate to calculate the enhancement factor. With a structure depth selected for optimized out-of-plane light–matter interaction for the WS$_2$-hBN-air-GaP system, we demonstrate strong PL enhancement due to the designed "air-GaP" configuration and additional yield owing to the underlying CIDBROM structure.

## Experiment

The stacking of isolated monolayer WS$_2$ onto multilayer hBN flakes suspended by the distinct air-GaP structures in the patterned substrate with in situ optical control took place under the microscope similar to the image sequence presented in **Figs. 2(a)-(f)**, shown for the transfer of a monolayer flake (real-color red glow under laser excitation, which is spectrally blocked here) onto the CIDBROM structure in **Fig. 1(c)**.

For the effective enhancement of the WS$_2$ PL at the A-exciton wavelength under green-light laser excitation, we calculated the expected enhancement factor for a suspended monolayer on an air-GaP structure as a function of the trench depth according to a multiple-reflection model (MRM, see Refs. [22] and Supporting Information for details). We determined a depth around 105 nm (for 10-nm-hBN-supported WS$_2$) to 135 nm (for directly suspended WS$_2$) to provide good conditions with theoretical enhancement factors above 100. In contrast, a 30-nm-thick hBN buffer would drastically reduce the enhancement effect by one order of magnitude, while the optimum structure depth needed to be 55 nm.

The desired bull's-eye-shaped microcavity structure with circular DBR (**Fig. 1(a-c)**) has been designed for high reflectivity in the structure's plane (see theoretical DBR spectrum in the Supporting Information) in accordance to the typical emission spectra of TMDCs. Here, the aim was to provide in-plane strong confinement conditions to the exciton emission wavelength of $WS_2$. For the etched in-plane-DBR structure with achieved air respectively GaP mean widths of 110 and 135 nm, the calculated stop band is shown in **Fig. 1(d)**. The tolerance with regard to the proportions is good enough to provide the desired reflectivity for our typical TMDC emission.

As photonic structures for visible light require patterning of materials in the nanoscale, FIB milling provides an alternative means of structuring to methods involving masking, resist-development and cleaning (lift-off) procedures. FIB etching can directly inscribe structures on the scale of tens of nanometers with reasonable aspect ratios, as the realized microcavity example in **Fig. 1(b, c)** with well-defined ring features shows.

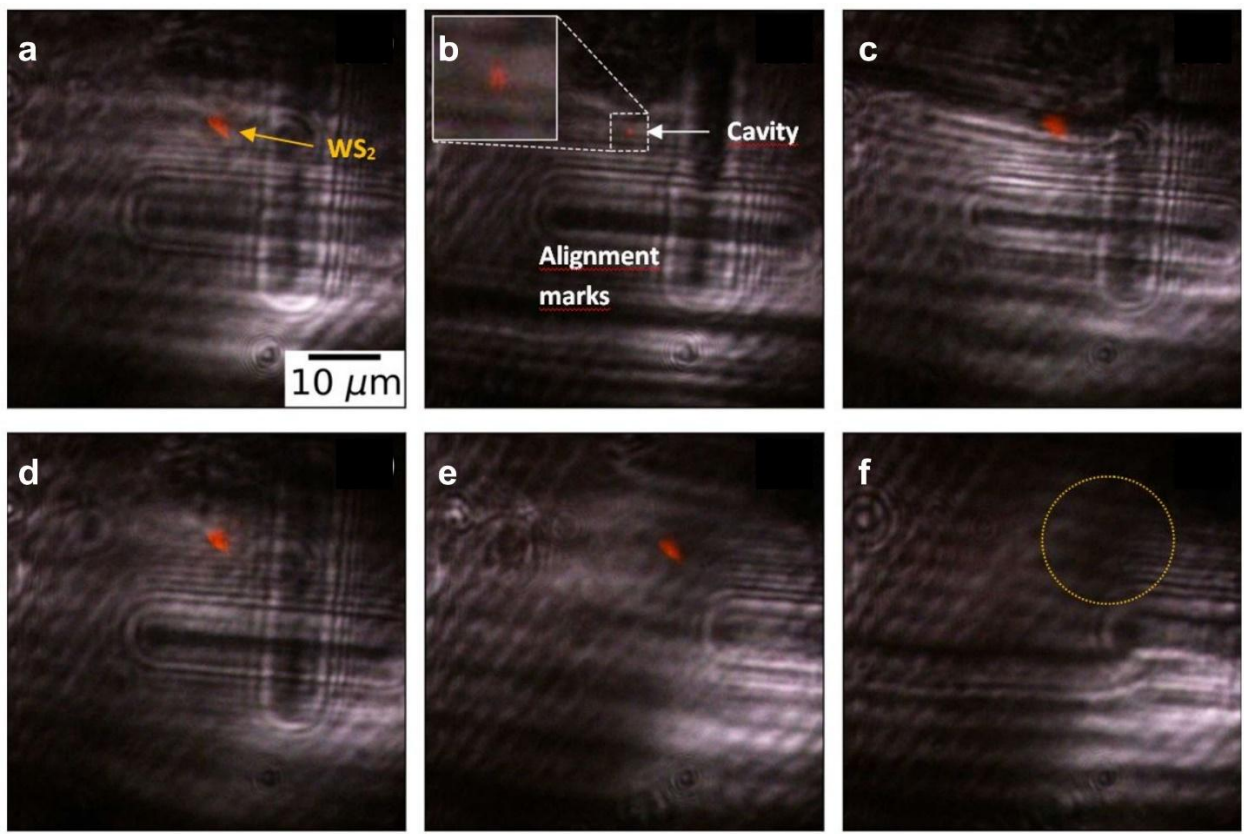

**Figure 2 | Room-temperature micro-photoluminescence images of a monolayer WS₂ flake on a polymer stamp sequentially brought in contact with the patterned CIDBROM sample.** *(a) Polymer gel film (stamp) with isolated monolayer WS₂ flake (micrograph, real-color image showing red PL) is touching the patterned GaP substrate in (b). While the GaP substrate quenches WS₂ PL, at the spot of the in-plane cavity, a circular PL spot remains (red glow). (c) The stamp partially in contact contains the lift-off flake during the stamp-removal process, showing PL again, with the flake focused in (d). (e) With the relative position of the CIDBROM changed with respect to the WS₂ flake location (c.f. aligment marks), PL disappears (f) when the flake is in contact with the bare substrate (encircled). Thus, the desired PL enhancement study requires an isolating layer between WS₂ and GaP, such as hBN.*

On the one hand, this approach can be particularly useful for patterning 2D-materials samples with high precision before or after layer-by-layer stacking, although one primary drawback of FIB milling is the unavoidable local contamination with—or incorporation of—the ion-beam particles into the structured material, here Gallium ions. However, for a GaP or GaAs sample, this is less critical than for other substrates. On the other hand, the sample should be conducting in order to enable the milling by the ions, which sets restrictions to the substrate choice.

Via FIB etching, the preparation of a number of optical structures such as photonic crystals [29–31], nanoring cavities [32], whispering-gallery-mode resonators [33] and microlenses [34] were demonstrated. In this work, a liquid metal ion source was used to create both a reference and a microcavity structure with feature depth on the order of 100 nanometers (130 nm for ML WS$_2$, 70 nm for hBN-buffered ML WS$_2$) in a GaP crystal.

Isolation of monolayers, stacking onto hBN and deposition onto the patterned GaP was performed similar to Ref. [35,36] using a gel film for a dry-stamping method. Here, a large enough hBN flake with optically homogeneous thickness was transferred onto the FIB-milled structures. Afterwards, exfoliated WS$_2$ monolayers were stamped on top of it to complete the stack. Micrograph images at different steps of this process are shown in the Supporting Information.

At first, we aimed at transferring an exfoliated and isolated flake of monolayer WS$_2$ onto the circular microcavity structure. Room-temperature micro-photoluminescence images of this flake on a polymer stamp and how it is sequentially brought in contact with the patterned CIDBROM sample are shown in **Figure 2**. While the monolayer on the polymer gel film (stamp) in the focal plane of the microscope can be clearly identified by its pronounced photoluminescence under irradiation with the green laser (filtered) in **(a)**, when touching the patterned GaP substrate in **(b)**, its PL is quenched by the GaP substrate. Only at the spot of the in-plane cavity, a circular PL spot remains (red glow). After the release from the surface, when the stamp is retreated partially, PL reemerges in the field of view focused on the monolayer in **(c, d)**. Repeating the deposition process with the relative position of the CIDBROM changed with respect to the WS$_2$ flake location (c.f. alignment marks, **(e)**), PL disappears **(f)** when the flake is in contact with the bare substrate (encircled area in the optical micrograph image). Thus, the desired PL enhancement study required an isolating layer between WS$_2$ and GaP, such as multilayer hBN in the thickness range of 10-20 nm, to provide minimum separation from the pattern—as the calculated enhancement factor significantly drops for increased hBN thickness—at sufficient isolation from the GaP surface. Nevertheless, this unexpected feature could be used for future on-chip monolayer optoelectronics on transparent GaP with site-controlled luminescence from patterned regions.

For this second approach, new and untouched structures in GaP prepared by FIB were used. The reference—a circular hole with the same diameter as the CIDBROM structure for the evaluation of the pattern's influence—and the microcavity are shown in **Fig. 3**. Both FIB-produced structures are shown in SEM images before WS$_2$-hBN deposition: the CIDBROM structure in GaP with about 110 nm depth in **(a)** in comparison to an analogously prepared reference circular-hole structure with about 130 nm depth in **(b)**. An inset indicates the underlying multiple-reflection model for optimized in- and out-coupling of light to and from the system at 532 nm and 615 nm, respectively.

**Figures 3(c)** and **(d)** show the measured topography for the individually prepared WS$_2$-hBN flakes on top of these two structures (microcavity and reference, respectively) in height-profile atomic-force-microscopy images. For comparability, the two air-GaP suspended 2D-material stacks needed to have comparable features. The monolayer areas on top of the buffering hBN multilayer flakes (13 nm for the cavity and 20 nm for the reference) are indicated by dashed lines, while dotted circles highlight the position of the respective patterns in the substrate. Cross-sectional height profiles for the substrate-to-hBN as well as hBN-to-monolayer steps are amended in the Supporting Information.

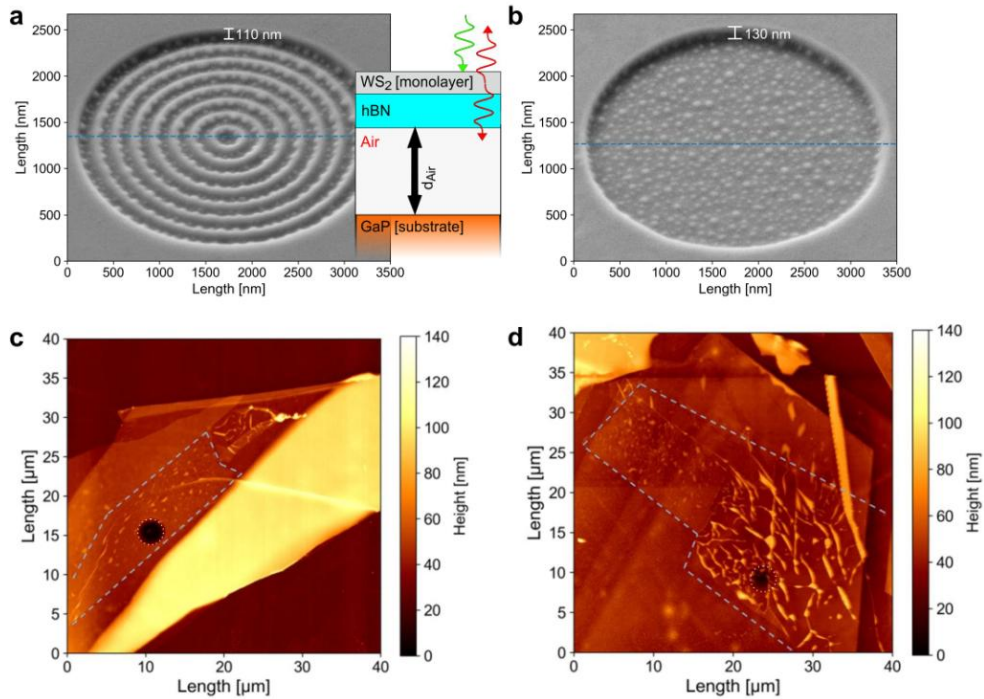

***Figure 3 | Microcavity as well as reference structure before and after WS₂-hBN deposition.*** *SEM image of a FIB-produced CIDBROM structure in GaP with about 110 nm depth (a) in comparison to an analogously prepared reference circular hole structure with about 130 nm depth (b). An inset indicates the underlying multiple-reflection model for optimized in- and out-coupling of light to and from the system at 532 nm and 615 nm, respectively. Corresponding height-profile atomic-force-microscopy images show the measured height-profile for the WS₂-hBN covered cavity (c) as well as reference structure (d). The monolayer areas on top of the buffering hBN multilayer flakes (13 nm and 20 nm, respectively) are indicated by dashed lines, while dotted circles highlight the position of the respective patterns in the substrate.*

## Results and Discussion

In order to obtain information about the effectiveness of the cavity, we investigated the local field distribution of the pump light with the help of scattering-type near-field optical microscopy (s-SNOM) on the Bragg cavity covered by the hBN/WS₂-monolayer and on the open cavity [37]. We performed near-field measurements at the PL excitation wavelength of 532 nm and—for comparison and as a reference—at an off-resonant out-of-stopband wavelength of 850 nm.

**Figure 4** shows the topography, **(a)**, and near-field signal, **(b)** and **(c)**, of the hBN/WS₂-monolayer-covered cavity, and for comparison the equivalent measurements on the open cavity, **(d-f)**, at the two measurement wavelengths at 532 nm (shown in green) and 850 nm (shown in red). While the grating structure is concealed in the topographical image in **Figure 4(a)** by the cover, its effect on the field enhancement on the in-coupled green light is quite distinct. The field map displays a prominent peak in the centre of the cavity at 532 nm **(b)**. In contrast to this, the field distribution for excitation at 850 nm shows a minimum in the centre **(c)**. This observation is reproduced for the equivalent measurements on the open Bragg cavity, suggesting that the field pattern is strongly influenced by the lateral grating structure and the desired field enhancement in the centre is obtained at the pump wavelength of 532 nm.

For a more detailed examination, we plot in **Figure 5** horizontal lineouts through the centre for each of the measurements presented in **Figure 4**. Vertical dashed lines indicate the borders and the centre of the grating structure. Firstly, the lineouts support the observation of a complementary behaviour, i.e., that the field is enhanced in the grating's centre at 532 nm, but reduced at 850 nm, outside of the stopband spectral region. Furthermore, while the near-field map at 850 nm follows the height of the grating, the near-field signal at 532 nm shows enhanced signals in the gaps between the grating structure for both the uncovered and covered sample. This pronounced difference may be influenced by the fact that s-SNOM measures two effects simultaneously, local electric fields and differences of the material properties (caused by corresponding variations of the scattering efficiency) [38]. The signal at 532 nm predominantly yields the field distribution with its maxima in the grating's gaps, while the signal at 850 nm is likely to have a strong spectroscopic material-sensing contribution. Moreover, the centre of the covered CIDBROM structure exhibits a stronger signal enhancement relative to the gaps between the rings.

These observations support the claim that the lateral periodic structure enhances the in-coupling of light at the excitation wavelengths of 532 nm spatially, and for the designed wavelength range. It should be noted that we repeated the s-SNOM measurement for the hole sample without any rings which is covered by hBN/WS$_2$-monolayer at 532 nm, where neither a ring pattern nor deterministic spatial enhancement in the s-SNOM data was observed (not shown here). Consequently, one is led to conclude that the PL of the WS$_2$ is additionally enhanced in the CIDBROM through an increased resonant in-coupling efficiency of the exciting light.

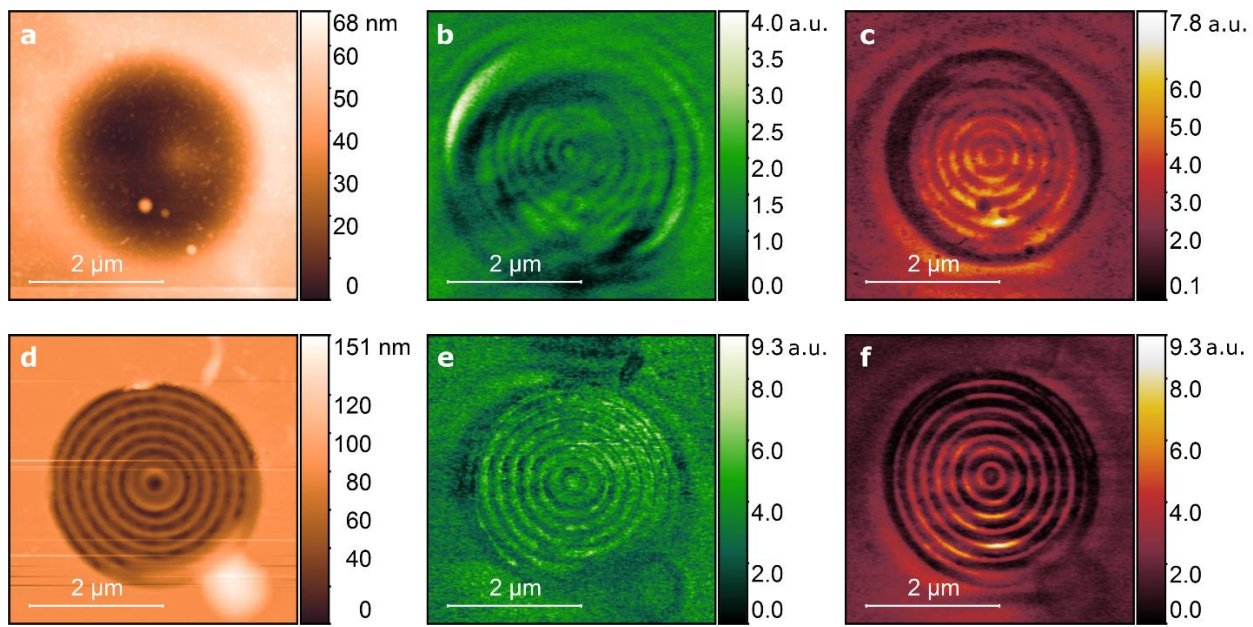

**Figure 4 | Topography and near-field measurements of monolayer-covered and open microcavity structure.** (a) Topography of WS$_2$-hBN covered cavity. (b) Near-field map of covered cavity at 532 nm. (c) Near-field map of covered cavity at 850 nm. (d) Topography of open microcavity. (e) Near-field map of open cavity at 532 nm. (f) Near-field map of open cavity at 850 nm.

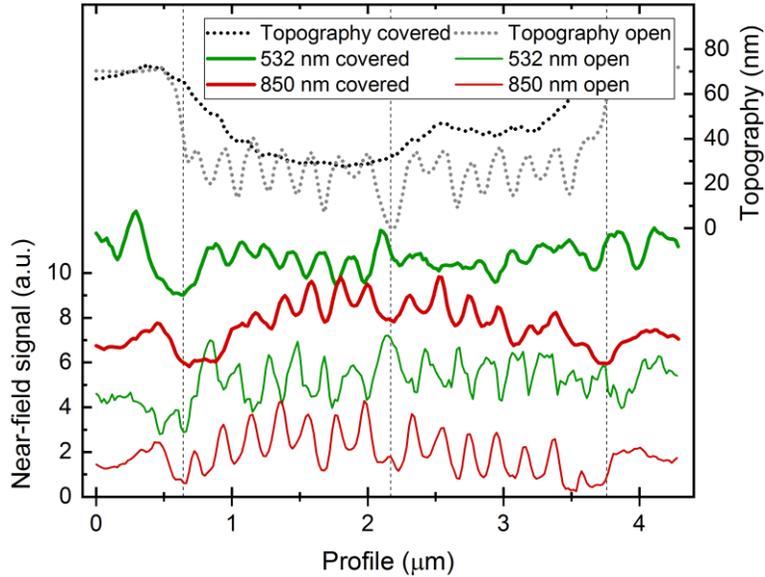

**Figure 5 | Horizontal line-outs taken from the AFM and s-SNOM measurement in Fig. 4.** *a) Dashed vertical lines mark the edges and centre of the microcavity structure. While the field maxima at 850 nm coincide with the topographical maxima of the Bragg cavity, the field distribution at 532 nm shows a complimentary behaviour. Most strikingly, one can observe a clear field maximum in the centre of the cavity for both measurements taken at 532 nm.*

Next, we analyzed the measured micro-PL of the 2D-stack-covered reference as well as microcavity structure. The emission from the monolayer areas was imaged by the spectrometer CCD (false-color picture) as shown in **Figure 6(a)**. A direct comparison of the signal levels from the CIDBROM to that of a similar monolayer area on top of the reference structure (**Figure 6(b)**) was possible due to the same acquisition conditions and settings. While a careful look reveals a small bright spot in the center of the cavity, no clear indication for a Purcell effect due to the cavity is seen in the spatial intensity profile. This is particularly overshadowed by the occurrence of the strong ring-shaped emission from the structure's off-center area attributed to efficient out-scattering of near-to-in-plane radiation by such in-plane grating. Moreover, the lack of resolution prevents reliable analysis of the spatial PL distribution.

The corresponding cross-sectional photoluminescence-intensity traces for both distinct structures in four distinguishable radial directions in 45° steps (see inset) are shown in **Figures 6(c)** and **(d)** in a direct comparison, while the intensity levels (counts per second per pixel) in **(a)** and **(b)** are in relation to each other. Clearly seen is the similar background level for the off-site micro-PL signal next to the structured area in the intensity profiles extracted. In contrast to this level, the intensity profile behaves very different for the two distinct structures. The maximum PL from the microcavity **(c)** reaches about 40k counts, while about 30k counts are obtained for the reference **(d)**. Here, a dark line visible in the inset of **(d)** and in the image in **(b)** causes an intensity drop, resulting in an unexpected non-symmetric profile for the reference. This artefact is attributed to a step in the hBN-buffer layer or to one of the many wrinkles in the transferred monolayer (c.f. AFM image), thus, two neighbouring regions on top of the relatively homogeneous hole structure have slightly different intensities.

To obtain spectral information for the relevant regions, reflection-contrast as well as PL spectra at on- and off-structure positions are shown for, both, cavity **(e)** and reference **(f)** structure in **Figure 6**, respectively. The reflection-contrast spectra, which show absorption features at room temperature, feature the A-exciton resonance on the structured areas with almost similar intensity dips. Here, the exciton signature for the reference case is slightly deeper. Only the off-site measurement for the stack near the reference hole seems to have contributions from both a monolayer as well as a bilayer section of the flake in the spectra, due to the blind way the off-site position was selected with closed real-space-projection aperture. Additionally, the high amount of wrinkles in this sheet might cause deviations from the ideal spectral lineshape. For the sake of comparability of the intensity counts, the real-space-image aperture in the detection path was closed to the projected size of the structure before acquiring both reflection and PL spectra consecutively. With the same setting, off-site monolayer spectra, off-site hBN-reference spectra and background measurements were performed, which allowed us to obtain very comparable reflection-contrast and PL information.

Ultimately, PL spectra show a clear enhancement of the exciton emission with respect to a completely unstructured GaP substrate for, both, the in-plane microcavity and the reference, while the ratio between off- and on-site signal in the two cases is 10.0 and 3.5, respectively.

Comparing the experimental ratio of cavity and reference enhancement $R_{\mathrm{exp}}$ with the theoretical one $R_{\mathrm{MRM}}$, that is

$$R_{\mathrm{exp}} = \frac{F_{\mathrm{Experiment,Resonator}}}{F_{\mathrm{Experiment,Reference}}} = \frac{10.0}{3.5} = 2.9,$$

$$R_{\mathrm{MRM}} = \frac{F_{\mathrm{MRM,Resonator}}}{F_{\mathrm{MRM,Reference}}} \cdot \frac{A_{\mathrm{eff,Reference}}}{A_{\mathrm{eff,Resonator}}} = \frac{68}{32} \cdot \frac{8.4\ \mu m^2}{5.7\ \mu m^2} = 3.1,$$

with $A_{\mathrm{eff}}$ and $F$ the corresponding effective suspended areas and enhancement factors, respectively, a good agreement regarding the MRM considerations can be seen. The systems parameters used for this calculation are summarized in **Table 1**. It follows that the spatially integrated PL enhancement was mainly caused by the interference effects perpendicular to the surface. However, in the spatially resolved PL measurement, an influence on the spatial intensity distribution on top of the CIDBROM structure could be observed. The locally higher intensity in the middle of the CIDBROM could even cause a Purcell amplification of the PL emission at this point, but this is not detectable with the given data.

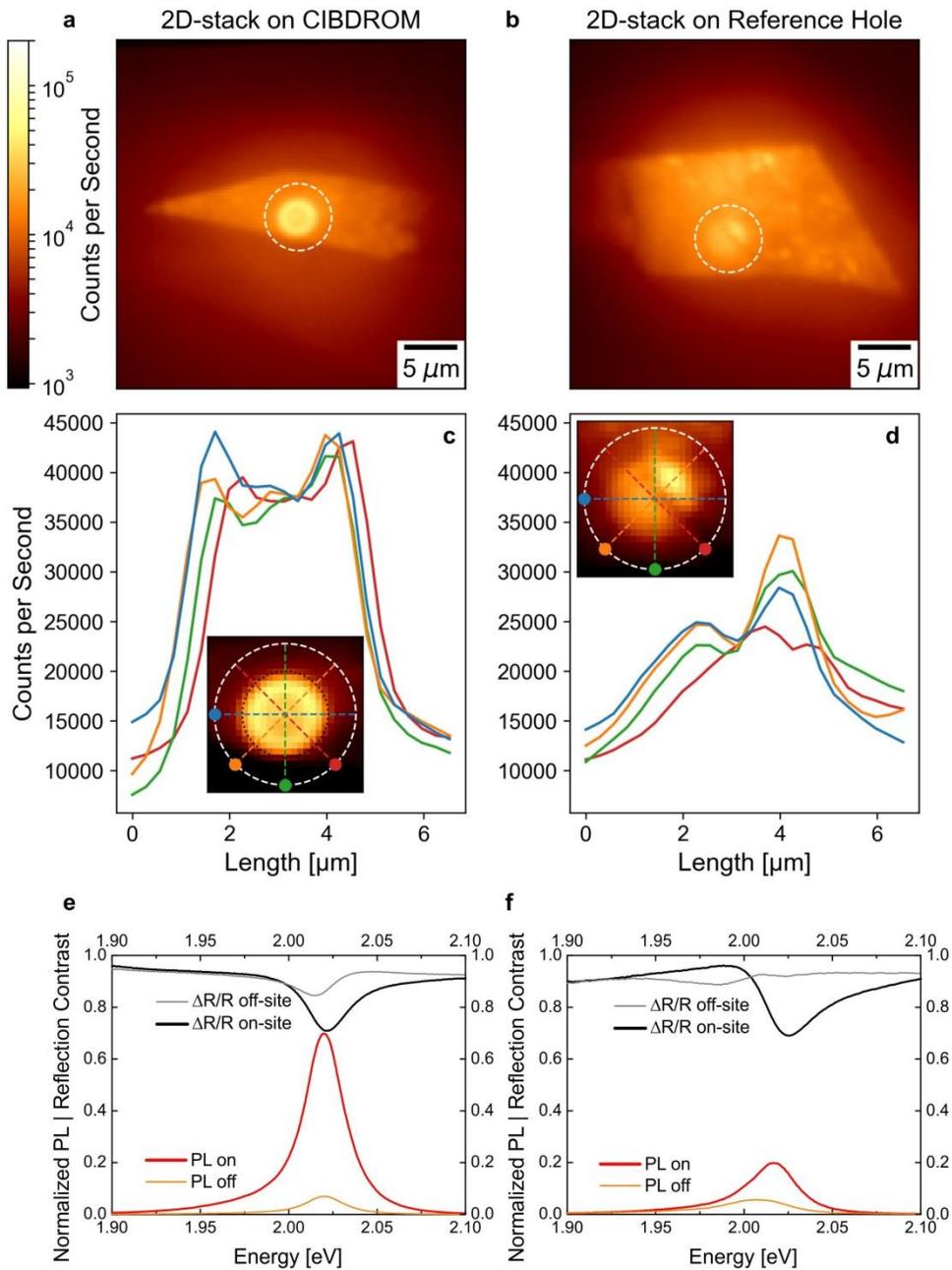

**Figure 6 | Photoluminescence enhancement obtained for a 2D-heterostructure-covered circular in-plane cavity in comparison to that of a reference hole.** *Here, micro-PL of the covered CIDBROM structure (a) with emission from the monolayer area imaged by the spectrometer CCD (false-color picture) is shown in direct comparison to a similar monolayer area on top of the reference structure (b). (c) and (d) show corresponding cross-sectional intensity traces (4 distinguishable radial directions in 45° steps, see inset) for both structures in a direct comparison, while the intensity levels (counts per second per pixel) in (a) and (b) are in relation to each other. The dark line visible in (b) and in the inset in (d) causes an intensity drop, resulting in an unexpected non-symmetric profile (artefact attributed to a step in the hBN buffer layer or a wrinkle in the monolayer). Reflection-contrast as well as PL spectra at on- and off-structure positions are shown for, both, cavity (e) and reference (f) structure, respectively.*

**Table 1 | Summary of the structural properties**

|  | Microcavity (on-site) | Microcavity (off-site) | Reference (on-site) | Reference (off-site) |
|---|---|---|---|---|
| Trench height (nm) | ~110 | 0 | ~130 | 0 |
| hBN thickness (nm) | 13 | 13 | 20 | 20 |
| 2D-layer dip (nm) | -40 | - | -25 | - |
| 2D height over GaP | 70 | | 105 | |
| Enhancement factor $F_{MRM}$ (theory) | 68 | | 32 | |
| Effective suspended area (µm²) $A_{eff}$ | 5.7 | | 8.4 | |
| PL peak position (eV) | 2.020 | 2.020 | 2.017 | 2.007 |
| FWHM (meV) | 27.1 | 26.5 | 33.3 | 46.4 |
| PL intensity ratio $I_{on}/I_{off} = F_{exp}$ | 10.0 | | 3.5 | |
| $F_{exp}/A_{eff}$ | 1.8 | | 0.4 | |

## Conclusions

In summary, we have presented a FIB-patterned structure capable of enhancing the WS$_2$ monolayer exciton's photoluminescence at room temperature by a factor of 10 with respect to an unstructured substrate using a combined structure scheme. A multiple-reflection model was used to optimize the etch depths of circular air-GaP structures for maximum constructive interference effects of the applied pump and expected emission light. In addition, a bull's-eye-shaped circular in-plane DBR-based optical microcavity was designed and realized, which showed clear differences in the enhancement properties in direct comparison to a reference circular hole structure, supported by photoluminescence, reflectance contrast and s-SNOM data. The effectiveness of this CIDBROM structure, as a model system for in-plane confinement and enhanced vertical PL extraction, is demonstrated with eyes towards future waveguide-coupled on-chip applications, e.g. for optical integrated circuitry and valleytronics, which requires further investigations on these structures as part of future endeavours.

## Methods

**FIB structuring:** Focussed-ion-beam milling took place directly on the GaP sample inside the evacuated chamber of a *Zeiss Auriya Dual Beam System* FIB device loaded with a Gallium gun. In situ imaging was possible using the integrated scanning-electron microscope.

**Optical measurements:** The measurement setup used in this study is similar to the one described in detail in Ref. [35], with the difference that an inverted commercial microscope (*OLYMPUS IX73*) was used. A schematic overview on the optical apparatus is given in the Supporting Information, with 2D-materials stamping section, light-source in-coupling as well as signal detection section. For the time-integrated PL measurements, the samples were irradiated quasi-resonantly at the B-exciton's energy with 532-nm continuous-wave laser light. White-light reflection spectra were obtained by a Tungsten lamp. The respective light beam is focused onto the sample using a 60x microscope objective. For optical control, a fiber-coupled Tungsten lamp, which was back-light illuminating the sample in the stacking section, and imaging CCD camera were used. For micrometer-precise spatial selection of the emission signal, an iris aperture was placed in the first available focal plane in the collected beam's path. For micro-PL measurements, a spectrometer with a 300 line (300-nm blaze) grating and a scientific EMCCD were used.

**AFM**: To determine the height steps of the multi-layer thin hBN buffer regions, on which the monolayers are placed, atomic-force microscope (AFM) images were acquired (see **Figure SI.4**). AFM was measured by a commercial device (*Agilent SPM 5500*) with Si-tip in tapping mode under ambient conditions at room temperature.

**s-SNOM:** The s-SNOM measurement set-up is a home-built system based on a modified AFM near-field microscope [39]. It uses a metallized AFM tip (*ARROW-NCPt, Nanoworld*) oscillating at $\frac{\Omega}{2\pi}$ = 260 kHz. Its apex is illuminated by a laser beam (P = 4 mW, p-polarized, diameter of the focal spot: 5 μm at a wavelength of either 532 nm or 850 nm. The light is focused onto the tip by a paraboloidal mirror (focal length f = 10 mm) and impinges onto the sample at an angle of about 60° relative to the surface normal. The tip is kept at a constant time-averaged distance of 200 nm from the surface while the sample is translated for raster-scanning via a XY-piezo-stage underneath it. The probe tip acts as a nano-antenna, scattering a part of the near-field wave into the far-field. The scattered light, non-linearly modulated at the frequency $\Omega$ of the tip oscillation, is collected by the paraboloidal mirror and detected using a silicon photodetector (*Thorlabs*) in the intensity detection scheme. The signal is demodulated at the third harmonic $\frac{3\Omega}{2\pi}$ thereby revealing the near-field signature. The spatial resolution is on the order of 20-50 nm [40]. s-SNOM measurements simultaneously yield the topography of the specimen via the AFM functionality and the spectroscopic near-field information via the tip-scattered light.

**Acknowledgement**
The authors acknowledge financial support by the German Research Foundation (DFG: SFB1083, RA2841/5-1), by the Philipps-Universität Marburg, by the Federal Ministry of Education and Research (BMBF) in the frame of the German Academic Exchange Service's (DAAD) program Strategic Partnerships and Thematic Networks, by the National Natural Science Foundation of China (No. 61635009), and by the Fundamental Research Funds for the Central Universities (No. 2018FZA5004). The authors thank G. Witte for AFM images and W. Wang for technical assistance with FIB etching. F. Walla acknowledges funding by the Friedrich-Ebert-Stiftung.

**Authors' contributions**
A.R.-I. initiated the study on in-plane microcavities as well as nanophotonic structures and guided the joint work together with W.F. The design and modeling of the structures was performed by O.M. who included the idea of utilizing the MRM for structure optimization. The implementation of the MRM and transfer-matrix model was done by F.W. The structures were processed by O.M., F.W., N.Y. and P.Q. and the experiments conducted together with L.M.S. AFM measurements were contributed by D.G., while s-SNOM experiments were performed by F.Walla, A.S. and H.G.R. The results were discussed with the support of all coauthors. The manuscript was written by O.M. and A.R.-I. with input from all co-authors.

**Corresponding author**
a.r-i@physik.uni-marburg.de

**Authors' statement/Competing interests**
The authors declare no conflict of interest

**Additional information**
Supplementary Information accompanies this paper

## Supporting Information

# Enhancement of the Monolayer WS₂ Exciton Photoluminescence with a 2D-Material/Air/GaP In-Plane Microcavity

Oliver Mey[1], Franziska Wall[1], Lorenz Maximilian Schneider[1], Darius Günder[1], Frederik Walla[2], Amin Soltani[2], Hartmut G. Roskos[2], Ni Yao[3], Peng Qing[3], Wei Fang[3], and Arash Rahimi-Iman*[1]

1) Department of Physics and Materials Sciences Center, Philipps-Universität Marburg, Marburg, 35032, Germany
2) Physikalisches Institut, Johann Wolfgang Goethe-Universität, 60438 Frankfurt am Main, Germany
3) State Key Laboratory of Modern Optical Instrumentation, College of Optical Science and Engineering, Zhejiang University, Hangzhou 310027, China

### Microcavity design considerations

In **Figure 1(d)** the calculated one-dimensional (1D) reflectance according to a transfer-matrix-method simulation of an ideal distributed-Bragg reflector (DBR) designed for monolayer TMDC resonances in the visible range is shown (black line), using λ-quarter layers of air and GaP. The design wavelength corresponds with the A-exciton line of WS₂. Here, the tolerance regarding high reflectivity with respect to feature sizes $d_{\text{Air}}$ and $d_{\text{GaP}}$ is visualized by the plots in **Figure SI.1**.

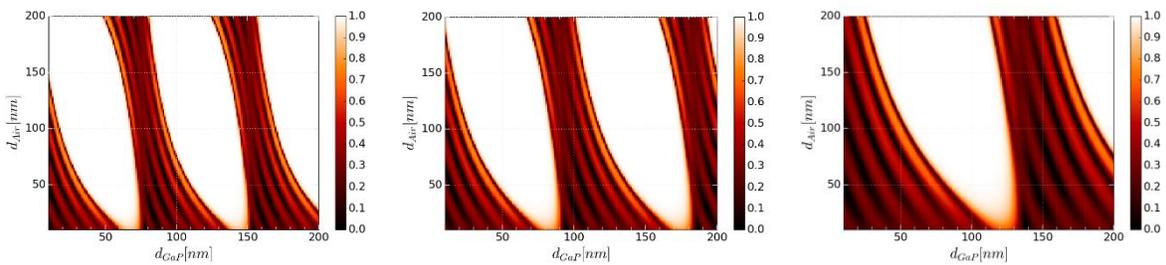

**Figure SI.1 | Reflectivity considerations:** Calculated reflectivity from the transfer-matrix method for air-GaP DBRs for different mean thicknesses of GaP and air features of the DBR, shown for the pump wavelength 532 nm, the design wavelength 615 nm, and an off-resonant wavelength of 850 nm. The realized CIDBROM structure of **Fig. 1** featured air and GaP rings of approximately 110 nm and 135 nm mean feature thicknesses, respectively. While for 532 and 615 nm, these feature sizes support a very high in-plane reflectivity, for 850 nm, the reflectivity is low and lateral confinement conditions are not provided.

## Microscope setup

The schematic representation of the (commercial inverted) optical microscope setup with excitation and detection pathways used for micro-photoluminescence and reflection-contrast measurements is shown in **Figure SI. 2**. The setup features a 2D-materials stacking section for site-controlled heterostructuring.

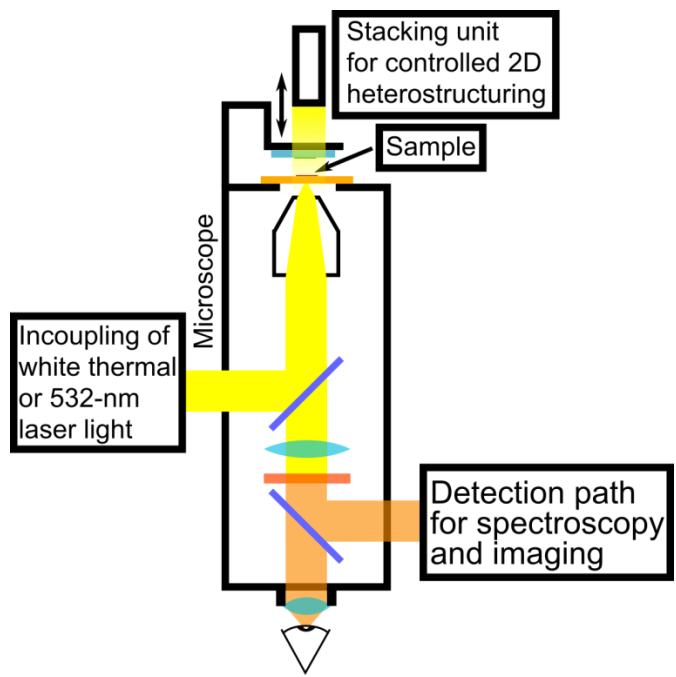

*Figure SI.2 | Scheme of the microscope setup.*

## Stacking of the 2D heterostructure

The intermediary steps towards the completion of an hBN-buffered WS$_2$ monolayer are shown as optical micrographs in **Figure SI.3**. Here, the first row shows the identified hBN flake with thin multi-layer area, on which the WS$_2$ flake with monolayer section, as presented in the second row, will be isolated. The resulting stack placed onto the patterned substrate is displayed in the last row. The two columns depict the 2D structures used for the reference (left) and the microcavity (right) structure with red-glowing monolayer parts due to laser excitation (532 nm, equal power level), respectively.

## Atomic-force microscopy

For the monolayer regions, one can clearly see quality differences in atomic-force microscope (AFM) images between the structure placed on the reference (left) and the counterpart placed on the CIDBROM structure (right in **Figure SI.4**). Here, imperfections in the 2D–2D interfaces and monolayer smoothness such as wrinkles and bubbles are evidenced as a consequence of the stacking process under ambient conditions using the dry-stamping method with polymer gel. The unavoidable bubbles could be further reduced by local accumulation of the sandwiched debris (e.g. water molecules) by annealing, while the wrinkles would remain. For the sake of comparability, no additional treatment was given to these structures, which were optically measured (at room temperature in ambient air) as is. The extracted cross sections for the monolayer-on-hBN (green dashed line) and hBN-on-GaP (blue dashed line) are shown in **Figure SI.4** below the micrographs with the same color code. The pronounced dips in the green curve correspond to the suspended 2D-structure areas bent into the air-GaP structures. This effect might be caused by the measurement tip of the AFM, or by a van-der-Waals-mediated pull of the 2D layers into the dips. The hBN buffers on the reference and microcavity were found to be 13 nm and 20 nm, respectively. The monolayer steps, confirmed by their optical properties, show a larger height step in AFM traces than 1 nm due to surface contamination with water molecules (from moisture) and other debris from the stacking process. This behaviour has been evidenced by us before and the expected height steps are commonly observed after thermal annealing in vacuum. However, such treatment was avoided here to reduce any risks of structure damage for these material stacks on our patterned air-GaP regions. It is important to note that µPL images and spectra clearly indicate monolayer emission regions.

## Multiple-reflection modelling of the optimum etching depth

The in- and out-coupling of light is optimized by taking interference effects into account. Here, a short overview about the idea of the multiple-reflection-model is given. A detailed derivation can be found in Ref. [22]. The in- and out-coupling efficiency corresponds to the electromagnetic field amplitude at the excitons' position $x$ for the incoming wavelength $\lambda_{in}$ and the emitted wavelength $\lambda_{out}$, respectively. To describe the field pattern for the wavelength $\lambda$ within a thin film structure, the propagation of the field amplitude $E$ within layer $i$ of thickness $d_i$ with a refractive index $n_i$ is modelled by a phase shift which is determined by the optical path $\beta_i$

$$E(x = 0) = E(x = d_i) \cdot e^{i \cdot \beta_i} \text{ with } \beta_i = \frac{2\pi n_i d_i}{\lambda}.$$

The WS₂ monolayer's thickness is taken as an effective optical thickness of 6.18 Å [23]. The behaviour at a single interface between layer $i$ and $j$ can be modelled by the Fresnel equations with a wavelengths dependent refractive index [23,41–44].

$$r_{ij} = \frac{n_i - n_j}{n_i + n_j} \text{ and } t_{ij} = \frac{2 \cdot n_i}{n_i + n_j}.$$

Here, the propagation direction of the light is assumed to be perpendicular to the layers' interface. If there is a stack of multiple thin films, the multiple-reflection-model takes all back-reflections by the structure into account and leads thereby to an effective reflection coefficient for an interface which consists of the reflection coefficient $a_1$ determined by the Fresnel equation as well as the sum over all backreflections $a_n$ (**Figure SI.5**). This summation can be solved with a geometric sum leading to the following equation

$$r_{12}^{eff} = a_1 + \sum_{n=0}^{\infty} a_2 \cdot \left(r_{21} \cdot e^{-2i\beta_2} \cdot r_{23}\right)^n = a_1 + \frac{a_2}{1 - r_{21} \cdot r_{23} \cdot e^{-2i\beta_2}}$$

If the structure consists of more than two thin layers above the substrate, it is necessary to calculate effective reflection coefficients for all interfaces recursively starting with the interface between the substrate and the first thin layer.

For the case of in-coupling, the field amplitude $E_{in}$ at position $x$ is calculated based on incoming light (**Figure SI.5**). The coefficient $b_1$ refers to the directly incoming light at position $x$., all other coefficients $b_n$ relate to the back-reflections between layer 0 and 1. This summation is split into odd and even coefficients $b_{2n+1}$ and $b_{2n+2}$ and can thereby also be solved by a geometric sum.

$$E_{in} = \sum_{n=1}^{\infty} b_n = \sum_{n=0}^{\infty} b_{2n+1} + b_{2n+2} = \frac{b_1 + b_2}{1 - r_{12}^{eff} \cdot r_{10} \cdot e^{-2i\beta_1}}.$$

For the out-coupling, the light path starts at the exciton's position. Our model is assumed to be one-dimensional. Therefore, the emitted light propagates only into two directions, up and down. With this, the emitted field amplitude can be described. The summation is solved in the same way as for $E_{in}$.

$$E_{out} = \frac{c_1 + c_2}{1 - r_{12}^{eff} \cdot r_{10} \cdot e^{-2i\beta_1}}$$

These field amplitudes, $E_{in}$ and $E_{out}$, are used to calculate the field intensities. Further, the exact position of the exciton is not taken into account in this model. It is assumed, that the excitons are equally distributed at all positions within the monolayer. Therefore, it is integrated over all positions $x$ in the final equation. Summed up, the enhancement factor $F$ describes the change in the measured intensity based on interference effects relative to some reference structure which is considered with the normalization factor $N$. The normalization is calculated by the following equation by inserting the reference structure as well as $N = 1$.

$$F = \frac{1}{N} \int_0^{d_1} |E_{in}(x) \cdot E_{out}(x)|^2 \, dx$$

It is important to mention that the enhancement factor is no absolute measure but a parameter for intensity ratios. Consequently, the normalization is as information essential together with the value for the factor.

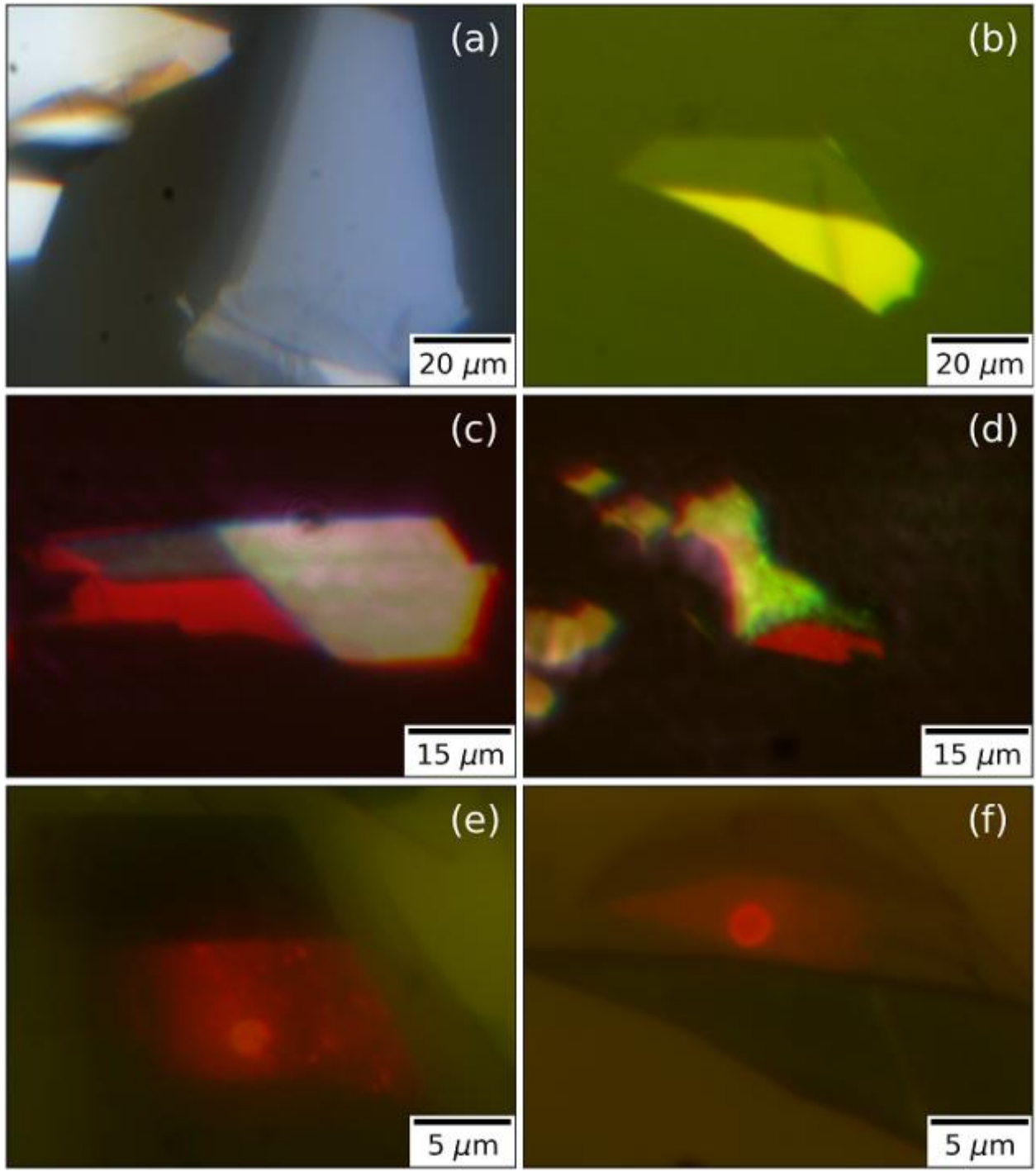

**Figure SI.3 | Sequence of the stacking process:** *The monolayer WS$_2$ (c and d) is placed on a multi-layer thin hBN (a and b) after its transfer to the patterned GaP reference (e) and microcavity (f) structures. The red glow of the layers under the microscope corresponds to monolayer emission excited by a 532 nm laser (filtered out).*

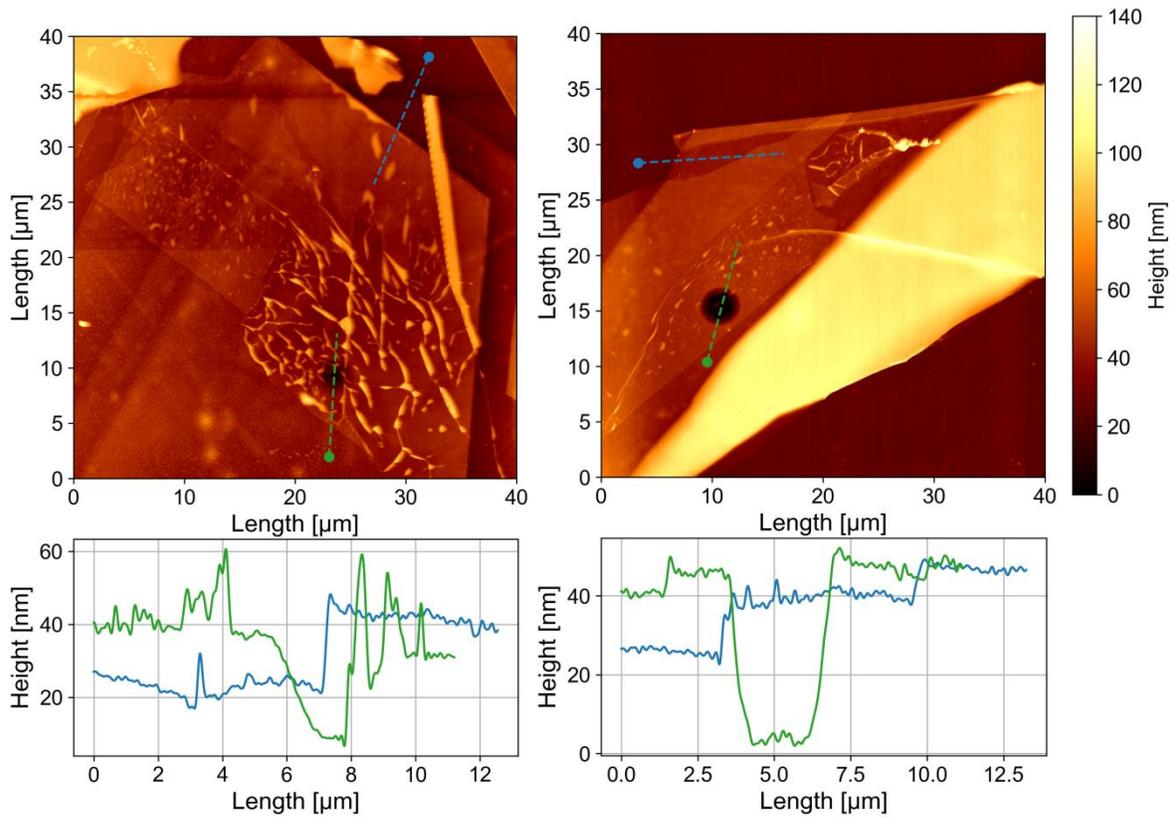

**Figure SI.4 | Raster-scanned height profiles of the 2D-structure covered patterned GaP substrates using atomic-force microscopy.** *The extracted cross sections for the monolayer-on-hBN (green dashed line) and hBN-on-GaP (blue dashed line) are shown below with the same color code for, both, the reference circular hole (left) and the in-plane microcavity (right).*

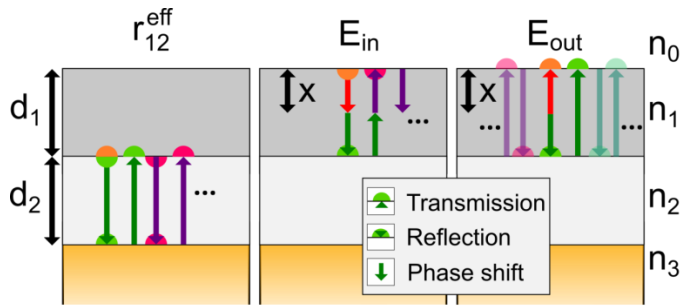

**Figure SI.5 | Schematics of the multiple-reflection model.**

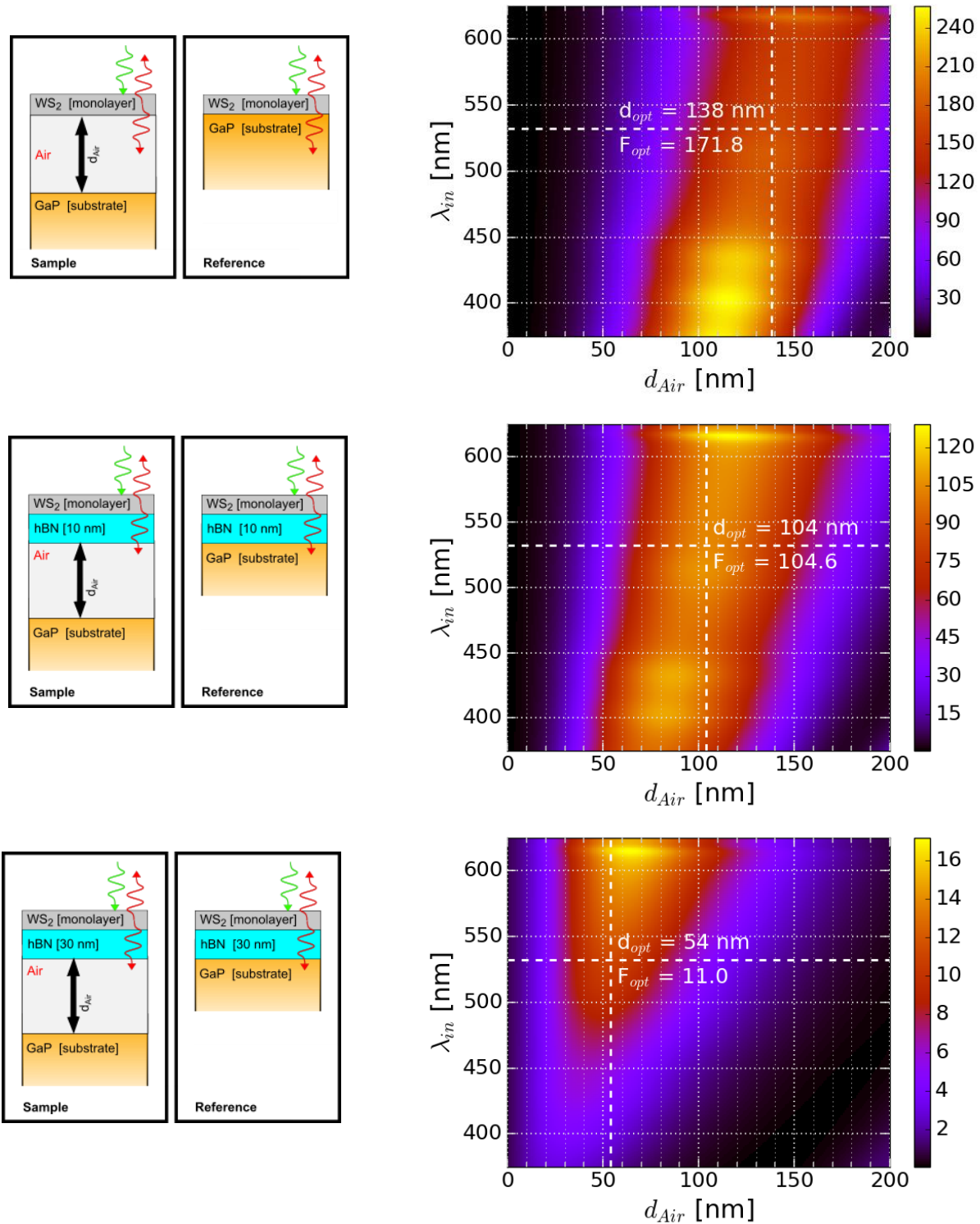

**Figure SI.6 | Enhancement factor for different structures:** On the left hand side, the sample as well as the reference structure are shown schematically. On the right hand side, the corresponding enhancement factor (sample divided by reference) is mapped for a varied incoming wavelength and thickness of the air gap. For the excitation-laser light ($\lambda_{in}$ = 532 nm), the optimal air-gap thickness ($d_{opt}$) to reach a maximal enhancement ($F_{opt}$) is marked by dashed lines.